\documentclass[pra,showpacs,twocolumn]{revtex4}
\usepackage{amssymb}
\usepackage{amsfonts}
\usepackage{amsmath}
\usepackage{graphicx}

\begin{document}
\title{Stochastic Representation of a Class of Non-Markovian Completely
Positive Evolutions}
\author{Adri\'{a}n A. Budini}
\affiliation{Max Planck Institute for the Physics of Complex Systems, N\"{o}thnitzer
Str. 38, 01187 Dresden, Germany}

\begin{abstract}
By modeling the interaction of an open quantum system with its environment
through a  natural generalization of the classical concept of continuous time
random walk, we derive and characterize a class of non-Markovian master
equations whose solution is a completely positive map.
The structure of these master equations is  associated with a random renewal
process where each event consist in the application of a
superoperator over a density matrix.
Strong non-exponential decay arise by choosing
different statistics of the renewal process.
As examples we analyze the stochastic and averaged dynamics of simple systems
that admit an analytical solution. The problem of positivity  in quantum master
equations induced by memory  effects $\mbox{[}$S.M. Barnett and S. Stenholm, 
Phys. Rev.~A {\bf 64}, 033808
(2001)$\mbox{]}$ is clarified in this context.
\end{abstract}


\maketitle

\section{Introduction}

From the beginning of quantum mechanics there existed alternative formalisms to
describe the dynamics of open quantum system. Besides the microscopic derivation
of quantum master equations, the theory of  quantum dynamical semigroups
\cite{alicki} introduced a strong constraint for the possible structure of a
given  Markovian master equation. As is well know, the
more general structure is given by the so called Kossakowski-Lindblad generator
\begin{equation}
\frac{d\rho \left( t\right) }{dt}=-i\left[ H,\rho \left( t\right)
\right] +
\frac{1}{\tau_{0}}{\cal L}_{0}\left[ \rho \left( t\right) \right]. \label{1-1}
\end{equation}
Here, $\rho(t)$ is the system density matrix, $H$ is the system Hamiltonian,
$\tau_{0}$ is the characteristic time scale of the irreversible dynamics
and
\begin{equation}
{\cal L}_{0}\left[ \bullet \right] =\sum_{\beta }
([V_{\beta },\bullet V_{\beta }^{\dagger }]+[V_{\beta }\bullet ,V_{\beta
}^{\dagger }]),  \label{zorra}
\end{equation}
where $\{V_{\beta}\}$ is a set of arbitrary operators. This structure arise
after demanding the Markovian property and the completely positive
condition (CPC). 
This last requisite is stronger than positivity.
It guarantees the right behavior of the solution map $\rho(0) \rightarrow 
\rho(t)$ after extending, with an identity, the original evolution to an 
ancillary and arbitrary Hilbert space \cite{alicki,nielsen}.

As a consequence of the Markovian or semigroup
condition, the evolution Eq.~(\ref{1-1}) is local in time. This fact, in
general, implies that the dynamics of the density matrix elements is
characterized through an exponential decay behavior.
Nevertheless, there exist many physical situations that must be
described in a quantum regime and whose characteristic decay behaviors are
different from an exponential decay.

Some relevant examples arise  in atomic and molecular systems subject
to the influence of environments with a highly structured spectral
density, where the theoretical modeling can be given in terms of
a few-modes spin-boson model \cite{gruebele} and in terms of random-matrix
theory\cite{wong}. In these situations, the characteristic decay of the system
dynamics present stretched exponential and power law behaviors. Other
examples are one dimensional quasiperiodic systems \cite{zhong} that develop a
non-Gaussian diffusion front, anomalous photon
counting statistics for blinking quantum dots \cite{barkai_dot}, many-spin
systems \cite{dobro}, fractional derivative master equations
\cite{kuznezov}, and structured reservoirs \cite{dalton}.

In all these  physical situations  the validity of the
approximations that allow a Markovian description break down. Therefore, its
dynamical description is outside of a Markovian Lindblad evolution.
Thus, there seems to be a gap between completely positive evolutions and
those with an anomalous decay behavior.


The main purpose of this paper is to establish the possibility of constructing a
class of evolution equations for the density matrix that satisfies the
CPC and that also lead to strong non-exponential 
decay. Our basic idea for the derivation of these equations consists in to model 
the
interaction of an open quantum system with its environment as a series of random
scattering events represented through the action of a superoperator over
the system density matrix, where the elapsed time between the successive
events corresponds to an arbitrary random renewal process \cite{feller}. This
stochastic dynamics can be seen as a natural generalization of the classical 
method 
of continuous time random walk \cite{montroll,metzler}, where
a particle at random times jumps instantaneously between the sites of a regular 
lattice.
In consequence we will name our starting
stochastic dynamics a continuous time quantum random walk (CTQRW).

We remark that the concept of quantum random walks is nowadays used in the
context of quantum information and quantum computation \cite{julia}. Our paper 
deals a different problem since here we are concerned with a
phenomenological description of anomalous irreversible processes in the context
of completely positive evolutions.

The dynamics that result from a CTQRW is non-Markovian
and can be written as a memory integral over a Lindblad superoperator [see
Eq.~(\ref{master})]. This kind of evolution was previously analyzed in Ref.
\cite{barnett} by Barnett and Stenholm, where was raised up  the possibility
of obtaining non physical solutions  from this non-Markovian evolution.
Contrarily to their final conclusion, here we will show that, as in a classical
context \cite{sokolov,barkai}, it is possible to use this kind
of equation as a phenomenological tool in the
description of open systems. Even more, we will see that the correct
behavior  of this equation is related with  the possibility of associating
to it a CTQRW.


The paper is organized as follows. In Section II we introduce the stochastic
dynamics and the corresponding evolution for the averaged density matrix. The
CPC and the relaxation to a stationary state are
characterized. In Section III we study some non trivial kernels that leads to a
telegraphic and a fractional equation. The dynamics induced by these evolutions
are analyzed through simples systems, as a two level system and a quantum
harmonic oscillator. The relation with the formalism of intrinsic decoherence is
also established. In section IV we give the conclusions.

\section{Continuous Time Quantum Random Walk}

The stochastic dynamics that define a CTQRW involve two central ingredients.
First, a completely
positive superoperator ${\cal E}[\bullet]$ which represent an
instantaneous disruptive intervention of the environment over the system of
interest. We will assume that it can be written in a sum representation
\cite{nielsen} as
\begin{equation}
{\cal E}[\rho ]=\sum_{i} C_{i}\rho C_{i}^{\dagger },\label{super}
\end{equation}
where the operators $C_{i}$ satisfies the closure condition
\begin{equation}
\sum_{i} C_{i}^{\dagger }  C_{i}=\text{I}. \label{uno}
\end{equation}
The second ingredient is a set of random time $t_{1}<t_{2}\cdots< t_{n}$
that define when the disruptive action occurs.
We will assume that this set is stationary and defined as a random renewal
process, i.e., it can be characterized through a waiting time distribution
$w(\tau)$ which gives the  probability density for the elapsed time interval
$\tau_{i}=t_{i}-t_{i-1}$ between two consecutive disruptive events.

We will work in an interaction representation with respect to the system
Hamiltonian and also assume that the unitary evolution commutates with the
superoperator ${\cal E}[\bullet]$. Thus, the average evolution of
the density matrix over the  realizations of the random times can be written in
the following way
\begin{equation}
\rho (t) =\sum_{n=0}^{\infty }P_{n}(t)\;{\cal E}^{n}[\rho(0)].\label{sol}
\end{equation}
Here, $P_{n}(t)$ defines the probability that $n$ applications of the
superoperator ${\cal E}[\rho]$ have occurred up to time $t$. This set
of probabilities is normalized as
\begin{equation}
\sum_{n=0}^{\infty }P_{n}(t)=1,
\end{equation}
and is defined through the expressions
\begin{equation}
P_{0}(t)=1-\int_{0}^{t}d\tau w(\tau),
\end{equation}
and
\begin{equation}
P_{n}(t)=\int_{0}^{t}d\tau w(t-\tau )P_{n-1}(\tau ) \label{Pn}.
\end{equation}
Note that $P_{0}(t)$ defines the survival probability, i.e., the probability of
having not any superoperator action up to time $t$.
Using recursively Eq.~(\ref{Pn}), from Eq.~(\ref{sol}) it is possible to express
the average density matrix as
\begin{equation}
\rho (t)=P_{0}(t)\rho (0)+\int_{0}^{t}d\tau w(t-\tau ){\cal E}[\rho (\tau )].
\label{integral}
\end{equation}
In order to obtain a differential equation for the evolution of $\rho(t)$
we follow the calculation in the Laplace domain. Denoting
$\tilde{f}(u)=\int_{0}^{\infty} dt
\exp[-ut] f(t)$, from Eq.~(\ref{integral}), we get
\begin{equation}
\tilde{\rho}(u)=\frac{1-\tilde{w}(u)}{u}\left\{ \frac{1}{\text{I}-\tilde{w}%
(u){\cal E}[\bullet ]}\right\} \rho (0).\label{laplace}
\end{equation}
where we have used $\tilde{P}_{0}(u)=[1-\tilde{w}(u)]/u$. Eq.~(\ref{laplace})
allows us to express $\rho(0)$ in terms of $\tilde{\rho}(u)$. Thus, it is
straightforward to get
\begin{equation}
u\tilde{\rho}(u)-\rho (0)=\tilde{K}(u){\cal L}[\tilde{\rho}(u)]\label{iraci}
\end{equation}
where we have defined
\begin{equation}
\tilde{K}(u)=\frac{u\tilde{w}(u)}{1-\tilde{w}(u)}, \label{kernel}
\end{equation}
and the superoperator
\begin{equation}
{\cal L}[\bullet ]={\cal E}[\bullet ]-\text{I}.\label{lindblad}
\end{equation}
Then, the time evolution of the average density matrix reads
\begin{equation}
\frac{d\rho (t)}{dt}=\int_{0}^{t}d\tau K(t-\tau ){\cal L}[\rho (\tau
)],\label{master}
\end{equation}
where the kernel $K(t)$ is defined through its Laplace transform
Eq.~(\ref{kernel}). This evolution, in general, is
non-Markovian, and by construction it is a
completely positive one. On the other hand, using the sum representation
Eq.~(\ref{super}) and the normalization condition
Eq.~(\ref{uno}) it is possible to write the superoperator
Eq.~(\ref{lindblad}) in a Lindblad form
\begin{equation}
{\cal L}[\bullet ]=\frac{1}{2}\sum_{i}\{[C_{i},\bullet C_{i}^{\dagger 
}]+[C_{i}\bullet
,C_{i}^{\dagger }]\}.
\end{equation}

{\it Random Superoperators}:  The previous results can be easily extended to
the case in which  the scattering superoperator, in each event, is
chosen over a set $\{{\cal E}_{a}[\bullet]\}$ with probability $P(a)da$.
Assuming  that this random selection is statistically independent of the set
of random times, the evolution is the same as in Eq.~(\ref{master})
with
\begin{equation}
{\cal L}[\bullet ]=\int_{-\infty }^{+\infty }daP(a){\cal E}_{a}[\bullet
]-\text{I}.\label{LA}
\end{equation}

{ \it Infinitesimal Transformations}: 
At this point, it is important to remark that in general an arbitrary Lindblad
structure, Eq.~(\ref{zorra}), can not be
associated with a completely positive superoperator ${\cal E}[\bullet]$ as in 
Eq.~(\ref{lindblad}).
This fact does not imply any limitation in our approach. In fact, 
an arbitrary Lindblad term ${\cal L}_{0}[\bullet]$
can be always associated to a completely positive superoperator of the form
\begin{equation}
{\cal E}_{0} [\rho]=\{\text{I}+[e^{\kappa {\cal L}_{0}}-\text{I}] \} \rho,
\label{infinitesimal}
\end{equation}
where $\kappa$ must be intended as a control parameter. Then, an arbitrary
Lindblad term can be introduced in Eq.~(\ref{master}) in the limit in which
simultaneously $\kappa \rightarrow 0$ and the number of events by unit of time
go to infinite, the last limit being controlled by the waiting time distribution
$w(t)$. We will exemplify this procedure along the next section.

\subsection{Completely Positive Condition}

As was mentioned previously, by construction the non-Markov evolution
Eq.~(\ref{master}) is a completely positive one. Nevertheless,
from a phenomenological point of view \cite{barnett} one is also interested 
to know which kind of arbitrary kernel $K_{d}(t)$
guarantee this condition.

The CPC is clearly satisfied if it is possible to 
associate to
the kernel $K_{d}(t)$ a well defined waiting distribution. Given an arbitrary
kernel, from the definition Eq~(\ref{kernel}), the associated waiting time
distribution is
\begin{equation}
\tilde{w_{d}}(u)=\frac{\tilde{K_{d}}(u)}{u+\tilde{K_{d}}(u)}=\frac{1}{u/\tilde{K_{d}}(u)+1}.
\label{waiting}
\end{equation}
This equation defines a positive waiting time distribution if and
only if $\tilde{w_{d}}(u)$ is a completely monotone (CM) function
\cite{feller}, i.e.
$\tilde{w_{d}}(0)>0$ and $(-1)^{n}\tilde{w_{d}}^{(n)}(u)\geqslant 0$, where
$\tilde{w_{d}}^{(n)}(u)$ denote the
n-derivative. After using
that $1/(u+1)$ is a completely monotone function, and that a function of the
type $f(g(u))$ is CM, if $f(s)$ is CM and if the function $g(u)$ is positive
and possesses a CM derivative \cite{feller},
the Laplace transform of the kernel $K_{d}(u)$ must satisfy
\begin{equation}
{\displaystyle{u \over \tilde{K_{d}}\left( u\right) }}%
\geqslant 0\text{ and }
{\displaystyle{d[u/\tilde{K_{d}}\left( u\right) ] \over du}}%
\text{ a CM function.} \label{condiciones}
\end{equation}
As in the classical case, these conditions allow us to classify the kernels in
safe and dangerous ones\cite{sokolov}. The secure ones, independently of the
particular structure of the superoperator ${\cal E}[\bullet]$, always admit a
stochastic interpretation in terms of a CTQRW. Therefore, they induce a
completely positive dynamics.
The dangerous ones do not admit a stochastic interpretation and in
consequence the CPC is not guaranteed. As we will
see in the next examples, in this last case the CPC 
depends on the particular structure of the superoperator ${\cal E}[\bullet]$.

\subsection{Integral Solution-Subordination Processes}

The solution of the evolution Eq.~(\ref{master}) can be written in an integral
form over the solution of a corresponding Markovian problem. In order to
demonstrate this affirmation, first we write Eq.~(\ref{iraci})
as
\begin{equation}
\tilde{\rho}(u)= \frac{1}{\text{u}-\tilde{K}%
(u){\cal L}[\bullet ]} \rho (0) . \label{hebe}
\end{equation}
Using the expression
\begin{equation}
\frac{1}{\text{u}-\tilde{K} (u){\cal L}[\bullet ]} =
\int_{0}^{\infty} d\tau^{'}
e^{ -\{ u-\tilde{K}(u) {\cal L}[\bullet] \} \tau^{'}},
\end{equation}
and after the change of variable $\tau= \tilde{K}(u) \tau^{'}$ it is possible to 
write
\begin{equation}
\frac{1}{\text{u}-\tilde{K} (u){\cal L}[\bullet ]} =
\int_{0}^{\infty} d\tau \tilde{P} (u,\tau)
e^{ {\cal L}[\bullet] \tau}, \label{eterna}
\end{equation}
where the function ${\tilde P}(u,\tau)$ is defined by
\begin{equation}
\tilde{P}\left( u,\tau \right) =\frac{1}{\tilde{K}\left( u\right) }\exp
[-\tau \frac{u}{\tilde{K}\left( u\right) }]. \label{mirna}
\end{equation}
Note that from this expression, after a Laplace transform in the second variable
$ \tau \to s$, it is possible to obtain ${\tilde P}(u,s)=1/[u+s
{\tilde K}(u)]$, which implies the equivalent definition
\begin{equation}
\frac{\partial P(t,\tau)}{\partial t}=-\int_{0}^{t}dt^{'} K(t-t^{'})
\frac{\partial
P(t^{'},\tau)}{\partial \tau}.
\end{equation}
Inserting Eq.~(\ref{eterna}) in Eq.~(\ref{hebe}), the  integral solution for the
density matrix reads
\begin{equation}
\rho \left( t\right) =\int_{0}^{\infty }d\tau P\left( t,\tau \right) \rho
^{(M)}\left( \tau \right),  \label{morena}
\end{equation}
where the density operator $\rho ^{(M)}(\tau) $
is the solution of  the Markovian evolution
\begin{equation}
\frac{d\rho ^{(M)}(\tau) }{d\tau}={\cal L}[\rho ^{(M)}(\tau)],
\end{equation}
subject to the initial condition $\rho(0)$, i.e.,
 $\rho ^{(M)}(\tau) =\exp[{\cal L} \tau] [\rho( 0)]$.

When the set of conditions Eq.~(\ref{condiciones}) is satisfied, from
Eq.~(\ref{mirna}) it is simple to demonstrate that the function $P(t,\tau)$
defines a probability distribution for the $\tau-$variable \cite{aclaracion},
i.e.,
\begin{equation}
P\left( t,\tau \right) \geqslant 0\text{\ \ \ and\ \ \ }\int_{0}^{\infty
}d\tau P\left( t,\tau \right) =1,  \label{claudia}
\end{equation}
where the normalization of $P\left( t,\tau \right) $ follows from
$\int_{0}^{\infty}d\tau P\left( u,\tau \right) =1/u$. This result, joint with
Eq.~(\ref{morena}), allows us to interpret the stochastic
evolution as a subordination process \cite{feller,sokolov}, where the
translation between the ``internal time" $\tau$ and the physical time $t$ is
given by the function $P(t,\tau)$. On the other hand, note that the positivity
of this probability function is equivalent to the CPC of the solution map.

\subsection{Relaxation to the stationary state}

Here we will analyze the relaxation of the density matrix to
a stationary state. With the aid of the integral solution Eq.~(\ref{morena}),
the characterization of this process is similar to that of classical
Fokker-Planck equations \cite{barkai}.
First, we note that the Markovian evolution of $\rho ^{(M)}(\tau)$  can be
always solved in a damping basis\cite{briegel} as
\begin{equation}
\rho ^{(M)}(\tau) =\sum_{\lambda }\check{c}_{\lambda }e^{-\lambda
\tau}P_{\lambda } , \label{relax}
\end{equation}
where $P_{\lambda}$ are the eigen-operators of the Lindblad term,
${\cal L}[P_{\lambda }]=\lambda P_{\lambda }$, and the expansion
coefficients are defined by $\check{c}_{\lambda }=$Tr$[%
\check{P}_{\lambda }\rho ^{(M)}\left( 0\right) ]$. The dual operators $%
\check{P}_{\lambda }$ satisfy the closure condition Tr$[\check{P}%
_{\lambda }P_{\lambda ^{%
{\acute{}}%
}}]=\lambda \delta _{\lambda \lambda ^{%
{\acute{}}%
}}$ and are defined through ${\check{{\cal L}}}[\check{P}_{\lambda
}]=\lambda \check{P}_{\lambda }$, where ${\check{{\cal L}}}[\bullet ]$ is
the dual superoperator of ${\cal L}[\bullet ]$ defined by Tr$\{A{\cal L}%
[\rho ]\}=$Tr$\{\rho {\check{{\cal L}}}[A]\}$\cite{alicki}. The expansion
Eq.~(\ref{relax}) allows us to write the solution of the non-Markov evolution
Eq.~(\ref{master}) in the form
\begin{equation}
\rho \left( t\right) =\sum_{\lambda }\check{c}_{\lambda }h_{\lambda
}(t)P_{\lambda }  \label{Relax2}
\end{equation}
where the functions $h_{\lambda }(t)$ are defined by
\begin{equation}
h_{\lambda }(t)=\int_{0}^{\infty }d\tau P\left( t,\tau \right) e^{-\lambda
\tau }.  \label{Relax3}
\end{equation}
In the Laplace domain this definition is equivalent to
\begin{equation}
\tilde{h}_{\lambda }\left( u\right) =\frac{1}{u+\lambda \tilde{K}\left(
u\right) } ,\label{Relax4}
\end{equation}
which also imply
\begin{equation}
\frac{d h_{\lambda }(t)}{dt} =-\lambda \int_{0}^{t} d\tau K(t-\tau)
h_{\lambda} (\tau).\label{Relax5}
\end{equation}
From these expressions it is simple to realize that if the Markovian solution
Eq.~(\ref{relax}) involves a null eigenvalue, the corresponding stationary state
maintains this status in the non-Markovian evolution.
Furthermore, the typical exponential decay of a Lindblad evolution is
translated to that of the characteristic functions $h_{\lambda}(t)$. On the 
other hand, 
due to the structure of the solution Eq.~(\ref{Relax2}), it is
clear that any set of relations between the relaxation rates of the Markovian
problem \cite{alicki,kimura} will be also present in the non Markov solution
[see Eqs.~(\ref{rateexponential})-(\ref{ratefractional})].

\section{Examples} \label{examples}

In this section we will analyze  different possible dynamics that arise after
choosing different memory kernels. Furthermore, we will work out some exact
solutions in simple systems.

\subsection{Markovian Dynamics}

By assuming an exponential waiting time distribution
\begin{equation}
w(t)=A_{1}e^{-A_{1}t},
\end{equation}
from Eq.~(\ref{kernel}) it is immediate to obtain
\begin{equation}
K(t)=A_{1} \delta(t).
\end{equation}
Thus, the evolution Eq.~(\ref{master}) reduces to a Markovian one.
In this case, it is also possible to obtain all the hierarchy of probabilities
$P_{n}(t)$, which read
\begin{equation}
P_{n}(t)=\frac{(A_{1}t)^{n}}{n!}e^{-A_{1} t}.
\end{equation}
This results imply that a Markovian Lindblad evolution
can be associated with a Poissonian statistics of the environment action.
This stochastic interpretation is also valid for arbitrary Lindblad terms
Eq.~(\ref{zorra}). In this case, the associated superoperator is given by
Eq.~(\ref{infinitesimal}) and it is necessary to take the limit $\kappa
\rightarrow 0$, $A_{1} \rightarrow \infty$ with $\kappa A_{1}=A_{1}^{'}$. Note
that this limit is well defined in the sense that the waiting time distribution
remains positive and normalized, i.e., $\int_{0}^{\infty} d\tau w(\tau)=1$.

\subsection{Exponential Kernel}

Now we will analyze the case of an exponential kernel
\begin{equation}
K\left( t\right) =A_{\epsilon} \exp [-\gamma t], \label{exponencial}
\end{equation}
where the units of $A_{\epsilon}$ are $sec^{-2}$.
By demanding the conditions Eq.~(\ref{condiciones}) it is possible to show that
this kernel is not a secure one, i.e., in general it is not possible to
associate a stochastic dynamics, and in consequence the CPC of the
solution map is not guaranteed. Nevertheless, note that in the double
limit, $
\gamma \rightarrow \infty $, $ A_{\epsilon} \rightarrow \infty
$, with $A_{\epsilon}/\gamma=A_{1}$ this kernel reduce to the previous case,
indicating a
possible region of parameters values where the kernel can be a secure one.
In order to see this fact, from Eq.~(\ref{waiting}), after Laplace transform,
we get
\begin{equation}
w(t)=2 A_{\epsilon} e^{-\gamma t/2} \frac{\sinh [\frac{1}{2} t
\sqrt{\gamma^{2}-4A_{\epsilon}}]}
{\sqrt{\gamma^{2}-4A_{\epsilon}}}. \label{exponential}
\end{equation}
This function, for $\gamma^{2}> 4A_{\epsilon}$ is a well defined waiting time
distribution which delimits the region of parameter values where the evolution 
is a secure one.

After differentiation of Eq.~(\ref{master}), the evolution of the density matrix
can be written as
\begin{equation}
\frac{d^{2}\rho \left( t\right) }{dt^{2}}+\gamma \frac{d\rho \left( t\right)
}{dt}=A_{\epsilon} {\cal L}[\rho \left( t\right) ]. \label{telegrafo}
\end{equation}
which is a kind of a telegraphic equation \cite{morse}.
This equation must be solved with the initial values $\rho \left( t\right) \mid
_{t=0}=\rho _{0}$ and $d\rho \left( t\right) /dt\mid _{t=0}=0.$ Then, under the
condition $\gamma^{2}> 4A_{\epsilon}$ this equation  provides an evolution whose
solution is a completely positive map.  In this case, the characteristic decay 
functions
$h_{\lambda}(t)$, Eq.~(\ref{Relax3}), results as
\begin{equation}
h\left( t,\Phi_{\lambda} \right) =e^{-\gamma t/2}\{\cosh
[\frac{t}{2}\Phi_{\lambda}
]+\frac{\gamma }{\Phi_{\lambda} }\sinh [\frac{t}{2}\Phi_{\lambda} ]\},
\label{dicotomico}
\end{equation}
where $\Phi_{\lambda}=\sqrt{\gamma^{2}-4 \lambda A_{\epsilon}}$.

We remark that the introduction of an arbitrary Lindblad term ${\cal L}_{0}[\bullet]$ in
Eq.~(\ref{telegrafo}) modifies drastically the previous positivity conditions.
In fact, this change requires the use of the
 superoperator Eq.~(\ref{infinitesimal}) and the double limit $\kappa \rightarrow 0$,
$A_{\epsilon} \rightarrow \infty$, with $\kappa A_{\epsilon}=A_{\epsilon}^{'}$.
Nevertheless, from Eq.~(\ref{exponential}), we note that the limit $A_{\epsilon}
\rightarrow \infty$ leads to a  waiting time distribution that always takes
negative values.
The positivity of $w(t)$ can only be recuperated in the limit $\gamma
\rightarrow \infty$. Nevertheless,  as we have commented previously, this extra
requirement  implies that the final dynamics converge to a Markovian ones.
Therefore, for infinitesimal superoperators there is no  region of parameter
values where the exponential kernel admits a stochastic interpretation. In consequence, the 
CPC of the solution map is unpredictable and must be checked 
for each particular case. This result characterizes and generalizes the results 
obtained in Ref.\cite{barnett}.

\subsection{Fractional Evolution}

Now we analyze a case of a sure kernel \cite{sokolov}. We assume
\begin{equation}
\tilde{K}\left( u\right) =A_{\alpha }u^{1-\alpha },\;\;\;\;\;0<\alpha \le 1,
\label{kernelfrac}
\end{equation}
where the units of $A_{\alpha}$ are  $1/sec^{\alpha}$. As is well known, this kind of kernel can
be related to a fractional derivative
operator \cite{metzler}. Thus, the density matrix evolution reads
\begin{equation}
\frac{d\rho \left( t\right) }{dt}=A_{\alpha \;0}D_{t}^{1-\alpha }{\cal L}%
\left[ \rho \left( t\right) \right] . \label{fraccionaria}
\end{equation}
The Riemann-Liouville fractional  operator is defined by
\begin{equation}
_{0}D_{t}^{1-\alpha }f(t)=\frac{1}{\Gamma (\alpha )}
\frac{d}{dt}\int_{0}^{t}dt^{\prime}
\frac{f(t^{\prime })}{(t-t^{\prime })^{1-\alpha }},
\end{equation}
where $\Gamma \left( x\right) $ is the Gamma function.
By using Eq.~(\ref{waiting}), the Laplace transform of the waiting
time distribution reads
\begin{equation}
\tilde{w}(u)=\frac{A_{\alpha }}{A_{\alpha}+u^{\alpha }}.
\end{equation}
Note that for $\alpha=1$ this expression reduces to the Laplace transform of
an exponential function. Furthermore, the condition $0<\alpha \le 1$ corresponds
to the values of $\alpha$ where ${\tilde w}(u)$ is a CM function, guaranteeing a well
defined waiting time distribution. In the time domain it reads
\begin{equation}
w(t)=\frac{A_{\alpha }}{t^{1-\alpha }}\sum_{n=0}^{\infty }\frac{(-A_{\alpha
}t^{\alpha })^{n}}{\Gamma \lbrack \alpha (n+1)]}. \label{serie}
\end{equation}
Thus, the case of fractional derivative provides a well defined evolution, Eq.~(\ref{fraccionaria}), whose solution is a completely positive map that admits a stochastic interpretation in terms of the waiting time distribution Eq.~(\ref{serie}). We remark that in this case, the average
time between successive applications, $\langle \tau \rangle=\int_{0}^{\infty}
 \tau w(\tau) d\tau$, is not defined.
As in the classical domain \cite{metzler}, this fact implies the absence of a
characteristic time scale and statistically it enables the presence of time
intervals of any magnitude.
On the other hand, we note that an arbitrary Lindblad superoperator can be
always introduced in
 Eq.~(\ref{fraccionaria}) in a secure way. In fact, the waiting time 
distribution
Eq.~(\ref{serie}) is well defined in the limit
 $\kappa \rightarrow 0$, $A_{\alpha} \rightarrow \infty$ with $\kappa
A_{\alpha}=A_{\alpha}^{'}$.

From Eqs.~(\ref{Relax4})-(\ref{kernelfrac}), the characteristic decay functions
$h_{\lambda}(t)$ read
\begin{equation}
h_{\lambda}(t)=E_{\alpha}[-\lambda A_{\alpha} t^{\alpha}].
\end{equation}
Here we have introduced the Mittag-Leffler function $E_{\alpha}(t)$
 which is defined through the series \cite{metzler}
\begin{equation}
E_{\alpha }\left[-A_{\alpha} t^{\alpha}\right] =\sum_{k=0}^{\infty
}\frac{(-A_{\alpha} t^{\alpha})^{k}}{\Gamma
\left(\alpha k+1\right) }.
\end{equation}
The short time regime of this function is governed by an stretched exponential
decay
\begin{equation}
\lim_{t\rightarrow 0}E_{\alpha }[- A_{\alpha} t^{\alpha }]\approx
e^{- A_{\alpha} t^{\alpha }},
\end{equation}
while the long time regime converges to a power law decay
\begin{equation}
\lim_{t\rightarrow \infty }E_{\alpha }[- A_{\alpha} t^{\alpha
}]\approx  \frac{1}{A_{\alpha} t^{\alpha} }.
\end{equation}
In this way, the fractional kernel allows us to introduce these
anomalous behaviors that clearly differ from the typical exponential decay of
a standard Lindblad equation.
Furthermore, this dynamics can be always
associated with a CTQRW characterized through the waiting time
distribution Eq.~(\ref{serie}).

\subsection{Short Time Regime}

An important aspect in the theory of open quantum systems is the
characterization of the irreversible dynamics at short times \cite{lu,budini}. Here we will
analyze this regime  through the linear entropy
$\delta(t)=1-Tr[\rho^{2}(t)]$. For simplicity, we will assume that at the
initial time the system is in a pure state, $\rho(0)=|\Psi\rangle\langle
\Psi|$. Defining the average
\begin{equation}
\langle\langle {\cal E} \rangle\rangle = \sum_{i} \langle C_{i}^{\dagger}
C_{i}\rangle-\langle
C_{i}^{\dagger}\rangle \langle C_{i} \rangle.
\end{equation}
where $\langle C \rangle = \langle\Psi|C|\Psi \rangle$, from
Eq.~(\ref{morena}), for the fractional case we get
\begin{equation}
\delta(t)\approx \frac{2  A_{\alpha} t^{\alpha}}{\Gamma (1+\alpha)}
\langle\langle {\cal E}
\rangle\rangle.
\end{equation}
while for the exponential case we get
\begin{equation}
\delta(t)\approx   A_{\epsilon} t^{2}\langle\langle {\cal E} \rangle\rangle.
\end{equation}
We note that for the Markovian case $(\alpha=1)$ the increase of entropy is
linear in time, while the exponential case present a slower quadratic
behavior.
On the other hand, the fractional case gives rise to the faster increase, whose
rate is not defined, i.e., it is infinite.
Nevertheless, as we will show in the next examples, in the long time regime
the fractional case induces the slower dynamical behavior.

\subsection{Two-Level System} \label{seccion}

Here we will analyze the non-Markovian dynamics of a two level system driven
by different superoperators and memory kernels.

\subsubsection{Depolarizing Reservoir}

First we will analyze the case of a depolarizing environment \cite{nielsen}.
Thus, we define the operators that appear in the sum
representation Eq.~(\ref{super}) as
\begin{equation}
C_{1}=\sqrt{p_{x}}\; \sigma _{x}\;\;\;\;C_{2}=\sqrt{p_{y}}\;\sigma
_{y}, \label{xy}
\end{equation}
where  $p_{x}+p_{y}=1$, and $\sigma_{x}$, $\sigma_{y}$ are the $x-y$ Pauli
matrixes. In order to simplify the final equations, from now on we will assume
$p_{x}=p_{y}=1/2$. In this case, the Lindblad superoperator ${\cal L}[\bullet]$
[Eq.~(\ref{lindblad})] reads
\begin{equation}
{\cal L} \left[ \bullet \right] =\frac{1}{4}([\sigma _{x},\bullet \sigma_{x}
]+[\sigma_{x}\bullet ,\sigma_{x} ]+%
[\sigma_{y} ,\bullet \sigma_{y}]+[\sigma_{y} \bullet ,\sigma_{y
}]).\label{infinito}
\end{equation}
This Lindblad generator corresponds to the interaction of a two level
system with a reservoir at infinite temperature. This fact can be clearly seen 
by expressing ${\cal L} [ \bullet]$ in terms of the lowering
and raising spin operators, $\sigma=(\sigma_{x}- i \sigma_{y})/2$,
$\sigma^{\dagger}=(\sigma_{x}+ i \sigma_{y})/2$.
We notice that assuming other values of $p_{x}$ and $p_{y}$, extra terms appear in
 Eq.~(\ref{infinito})  that do not modify the
infinite temperature property of the Lindblad superoperator.

{\it Exponential Kernel}: By denoting the density matrix $\rho (t)$
in the basis of the eigenvalues of $\sigma_{z}$ as
\begin{eqnarray}
\rho(t) =\left(
\begin{array}{cc}
P_{+}(t) & C_{+}(t) \\
C_{-}(t) & P_{-}(t)
\end{array}
\right),
\end{eqnarray}
from Eq.~(\ref{telegrafo}), the evolution of the upper and lower levels reads
\begin{equation}
\frac{d^{2}P_{\pm }(t)}{dt^{2}}+\gamma \frac{dP_{\pm }(t)}{dt}= A_{\epsilon}
\lbrack \pm  P_{-}(t)\mp  P_{+}(t)],
\end{equation}
while the coherences evolve as
\begin{equation}
\frac{d^{2}C_{\pm }(t)}{dt^{2}}+\gamma \frac{dC_{\pm }(t)}{dt}=- A_{\epsilon}
C_{\pm
}(t).
\end{equation}
The solution of these equations are
\begin{equation}
P_{\pm }(t)=P_{\pm }^{eq}+(P_{\pm }(0)-P_{\pm }^{eq})h(t,\Phi
_{pop}),\label{popular}
\end{equation}
with $P_{\pm }^{eq}=1/2$, and
\begin{equation}
C_{\pm }(t)=C_{\pm }(0)h(t,\Phi _{coh}), \label{coherente}
\end{equation}
where the function $h(t,\Phi)$ was defined in Eq.~(\ref{dicotomico}) and
\begin{equation}
\Phi _{pop} =\sqrt{\gamma^{2}-8  A_{\epsilon}},\;\;\;\;\;\;\;\;\;\;\;
\Phi_{coh}=\sqrt{\gamma^{2}-4 A_{\epsilon}}.
\end{equation}
In the Markovian limit $\gamma \rightarrow \infty $, $A_{\epsilon} \rightarrow
\infty$ with $A_{\epsilon}/\gamma=A_{1}$ we get the well know Markovian
results
$h\left( t,\Phi
\right) =\exp [-\Phi t],$ with $\Phi
_{pop}=2 A_{1}$ and $\Phi _{coh}= A_{1}.$

From our previous results, we know that under the condition $\gamma >4
A_{\epsilon}$ the dynamics must be a completely positive one and that for
$\gamma <4 A_{\epsilon}$ this is not  guaranteed.
Here, we will check  these conclusions for this simple model.
By using the property $|h(t,\Phi)|\le 1$, from Eq.~(\ref{popular}) it is
possible to conclude that for any value of the parameter $\gamma$ and
$A_{\epsilon}$, at all times the populations satisfies $P_{\pm}(t) \ge 0$.
On the other hand, the determinant $d(t)$ of $\rho(t)$, for
any parameter values, satisfies the inequality
\begin{eqnarray}
d(t)&=&\bigg\{\frac{1}{4}+\left(P_{+}(0)-\frac{1}{2}\right)\left(P_{-}(0)-\frac{1}{2}
\right) h^{2}(t,\Phi_{pop}) \nonumber
\\&-& C_{+}(0) C_{-}(0) h^{2}(t,\Phi_{coh}) \bigg \}\ge 0.
\end{eqnarray}
In consequence, independently of the values of $\gamma$ and $A_{\epsilon}$, the
density matrix is always positive. We remark that this result does not imply
that the solution map $\rho(0) \rightarrow \rho(t)$ is a completely positive
one. By writing the solution in the sum representation
\begin{equation}
\rho(t)=g_{I}(t) \;\rho(0)+\sum_{j=x,y,z} g_{j}(t)\; \sigma_{j} \rho(0)
\sigma_{j},\label{mapa}
\end{equation}
where
\begin{eqnarray}
g_{I}(t)&=&\frac{1}{2} \left[\frac{ 1+h(t,\Phi_{pop})}{2}+h(t,\Phi_{coh})
\right],
\\ g_{x}(t)&=& g_{y}(t)=\frac{ 1-h(t,\Phi_{pop})}{4},
\\ g_{z}(t)&=&\frac{1}{2} \left[\frac{ 1+h(t,\Phi_{pop})}{2}-h(t,\Phi_{coh})
\right],
\end{eqnarray}
the CPC is equivalent to the conditions
$g_{I}(t) \ge 0$, and  $g_{j}(t) \ge 0~\forall j$, for all times.
For $\gamma^{2} \ge 4 A_{\epsilon}$ these inequalities are satisfied.
On the other hand, for $\gamma^{2} \le 4 A_{\epsilon}$,
while the functions $g_{x}(t)$ and $g_{y}(t)$ are still positive,  the
functions $g_{I}(t)$ and $g_{z}(t)$ take
negative values, which imply that the map $\rho(0) \rightarrow \rho(t)$ is not a
completely positive ones.
Note that in this situation, the map Eq.~(\ref{mapa}) can be written
as a difference of two completely positive maps. This fact agrees with the
general results of Ref. \cite{yu}, where it was demonstrated that any positive
map can be written as a difference of two completely positive ones.
\begin{figure}[tbp]
\includegraphics[bb=90 140 692 582,angle=0,width=8.cm]{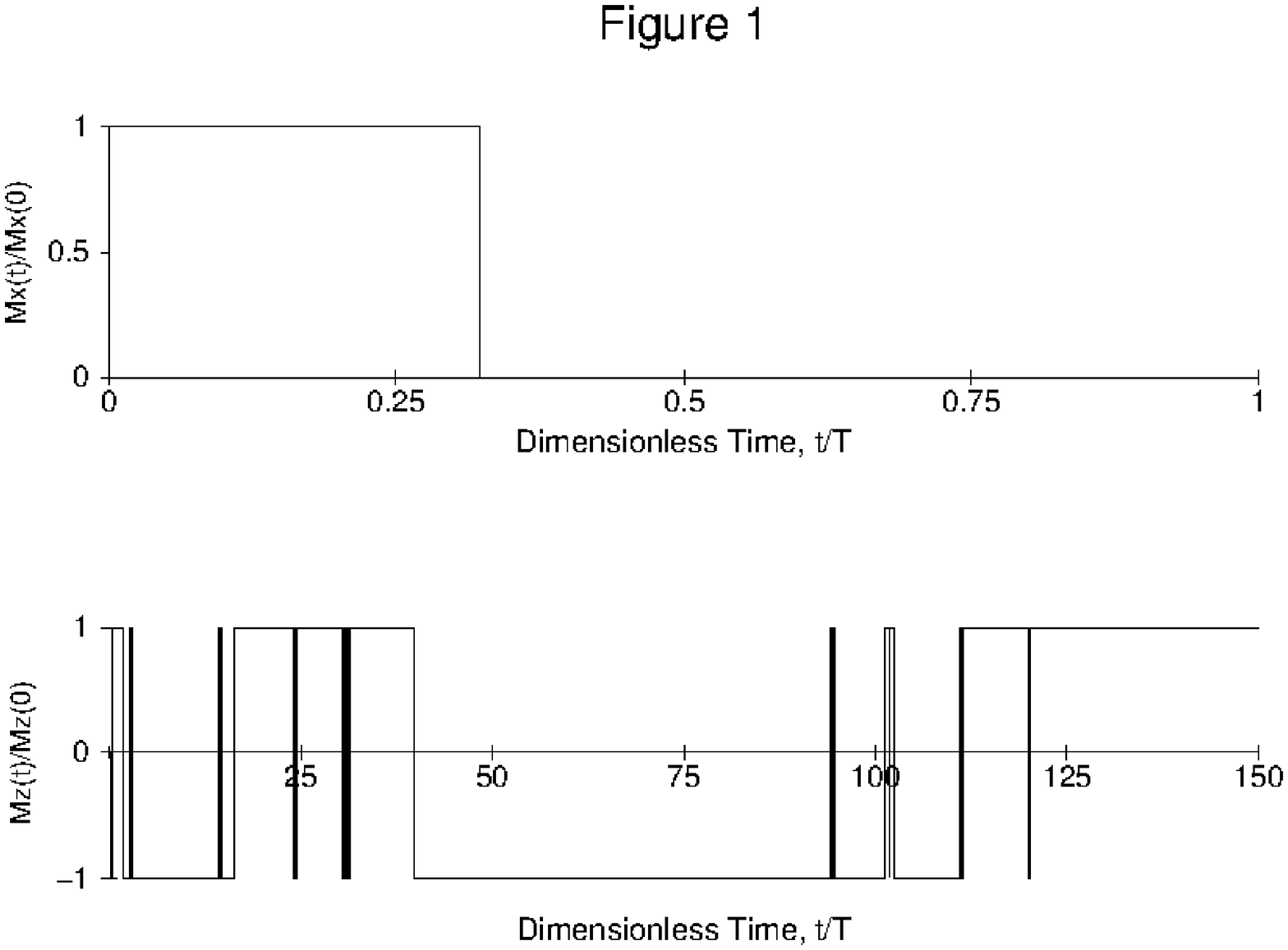}
\caption{Stochastic realizations for the CTQRW defined by the
depolarizing operators Eq.~(\ref{xy})
and the fractional waiting time distribution Eq.~(\ref{serie}). The
graphs correspond to the quantum average of the Pauli matrixes,
$M_{j}(t)=Tr[\rho(t) \sigma_{j}]$, $j=x,y,z$. The realization for the normalized
average $M_{y}(t)/M_{y}(0)$ is equal to that of the $x$-direction. 
The parameters were chosen as  $p_{x}=p_{y}=0.5$ and
 $\alpha=0.5$, $A_{\alpha}=1/\sqrt{2}~sec^{-1/2}$, T=$A_{\alpha}^{-1/\alpha}.$}
\label{realizations}
\end{figure}

{\it Fractional kernel}: Now we analyze the dynamics of the the two level
system in the case of the fractional kernel Eq.~(\ref{fraccionaria}). For the
evolution of the populations we get
\begin{equation}
\frac{dP_{\pm }(t)}{dt}= A_{\alpha \;0}D_{t}^{1-\alpha }
 [\pm P_{-}(t)\mp  P_{+}(t)],
\end{equation}
and the evolution of the coherence is
\begin{equation}
\frac{dC_{\pm }(t)}{dt}=- A_{\alpha \;0}D_{t}^{1-\alpha } C_{\pm
}(t).
\end{equation}
The solutions of these equations are
\begin{equation}
P_{\pm }(t)=P_{\pm }^{eq}+(P_{\pm }(0)-P_{\pm }^{eq})
E_{\alpha}[-\Phi_{pop}^{(\alpha)}
t^{\alpha}] \label{popfrac},
\end{equation}
and
\begin{equation}
C_{\pm }(t)=C_{\pm }(0)E_{\alpha}[- \Phi_{coh}^{(\alpha)}
t^{\alpha}], \label{coherfrac}
\end{equation}
where
\begin{equation}
\Phi_{pop}^{(\alpha)}=2 A_{\alpha},\;\;\;\;\;\
\Phi_{coh}^{(\alpha)}=A_{\alpha}.
\end{equation}
These expressions provide a completely positive map that admit a stochastic
interpretation in terms of its associated CTQRW.
%
%
%
%
In Fig.~(\ref{realizations}) we have implemented a numerical simulation of this quantum
 stochastic process. We show a set of realizations for the quantum averages
 of the Pauli matrixes, $M_{j}(t)=Tr[\rho(t) \sigma_{j}]$, $j=x,y,z$. 
 After the first application of the depolarizing superoperator, Eq.~(\ref{xy}), 
the  normalized values of $M_{x}(t)$ and $M_{y}(t)$ go to zero, remaining in this value at all 
 subsequent times. On the other hand, $M_{z}(t)/M_{z}(0)$
 oscillates between $\pm 1$ after each scattering event.
 A notable property of
these realizations is the absence of a characteristic time scale  both for the first event and  for the elapsed time between any successive events. 
This fact is a consequence of the power law 
decay of the waiting time distribution $w(t)$, Eq.~(\ref{serie}).
The absence of any time scale can be seen in the realization of $M_{z}(t)/M_{z}(0)$ where 
it is evident the presence of time intervals of any magnitude.

\begin{figure}[tbp]
\includegraphics[bb=80 77 690 581,angle=0,width=8.cm]{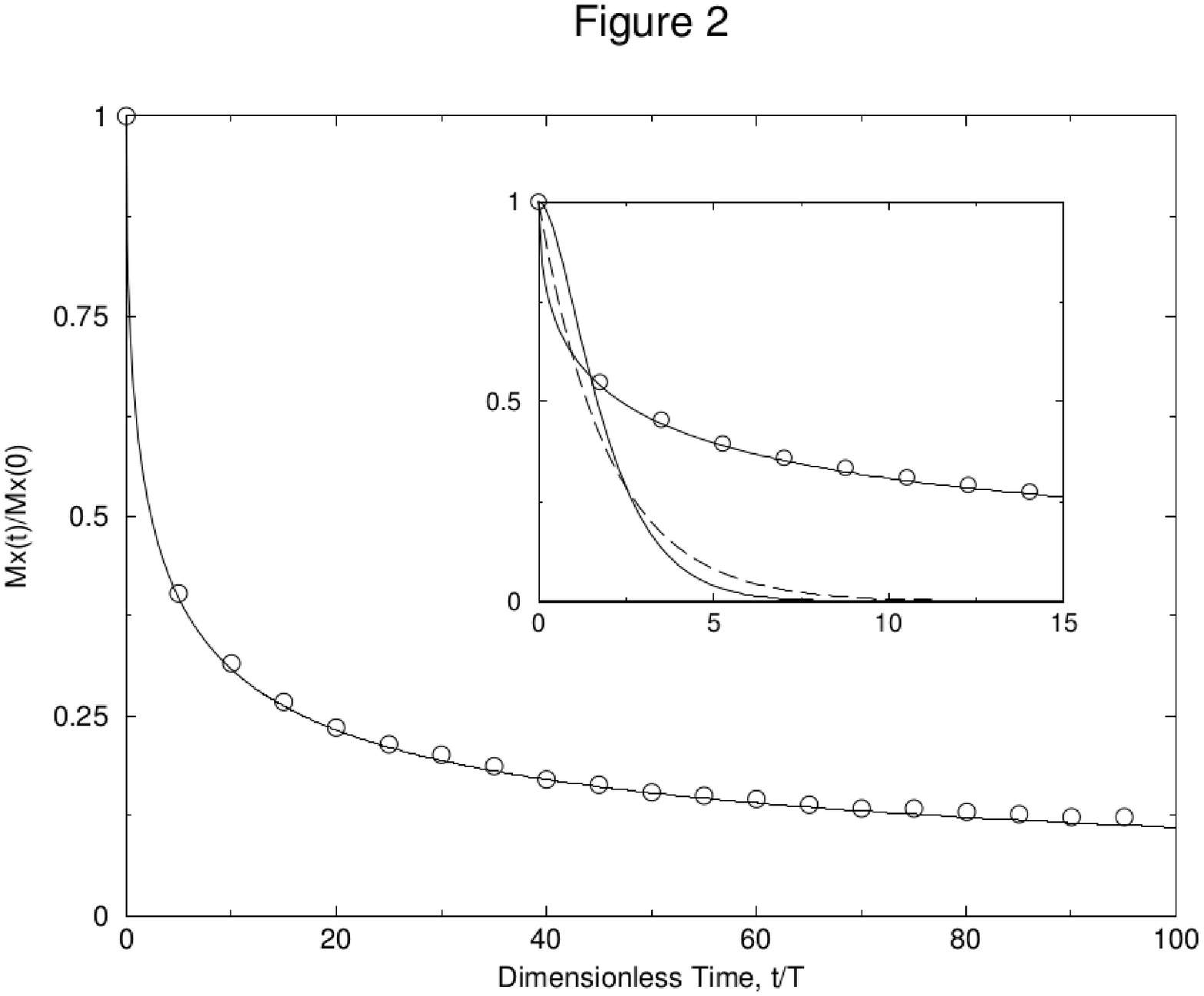}
\caption{Theoretical result (full line) and average over $10^{4}$ realizations
(circles) for $M_{x}(t)$. The inset shows the short time regime together with the
theoretical results for the
Markovian evolution (dashed line) with $A_{1}=0.5~sec^{-1}$, and  the
exponential kernel (full line) with $ \gamma=2~sec^{-1}$,
$ A_{\epsilon}=1~sec^{-2}$.}
\label{coherenceaverage}
\end{figure}

In Fig.~(\ref{coherenceaverage}) we show  the corresponding average over
$10^{4}$ realizations together with the analytical result for $M_{x}(t)$.
We have taken $\alpha=1/2$, which allows to use the equivalent
expression $E_{1/2}[A_{\alpha}t^{1/2}]= \exp[A_{\alpha}^{2} t]\;
\rm{erfc}[A_{\alpha} t^{1/2}]$ \cite{metzler}.
In the inset we compare the decay behavior induced by the different
kernels. Here, the stretched exponential decay at short times
and the power law behavior at long times are evident. In order to be able to
compare the different
time decay scales induced by each kernel, in all figure of the paper we
take $\{A_{1}=A_{\alpha}^{1/\alpha}=A_{\epsilon}/\gamma\} \equiv T^{-1}$, which
define the dimensionless time scale $t/T$.


{\it Linear entropy}: The linear entropy $\delta(t)$ can be used as a probe of
the density matrix positivity. In fact, in a two
dimensional Hilbert space, the positivity condition $\rho(t) \ge 0$ is
equivalent to the inequality  $0\leq \delta (t)\leq 1$. This means
that if one of the two eigenvalues of $\rho(t)$ is negative, them $\delta(t)<
0$. Furthermore, the dynamical behaviors induced by each kernel can be
shown in a transparent way through this object.

\begin{figure}[tbp]
\includegraphics[bb=80 77 690 581,angle=0,width=8.cm]{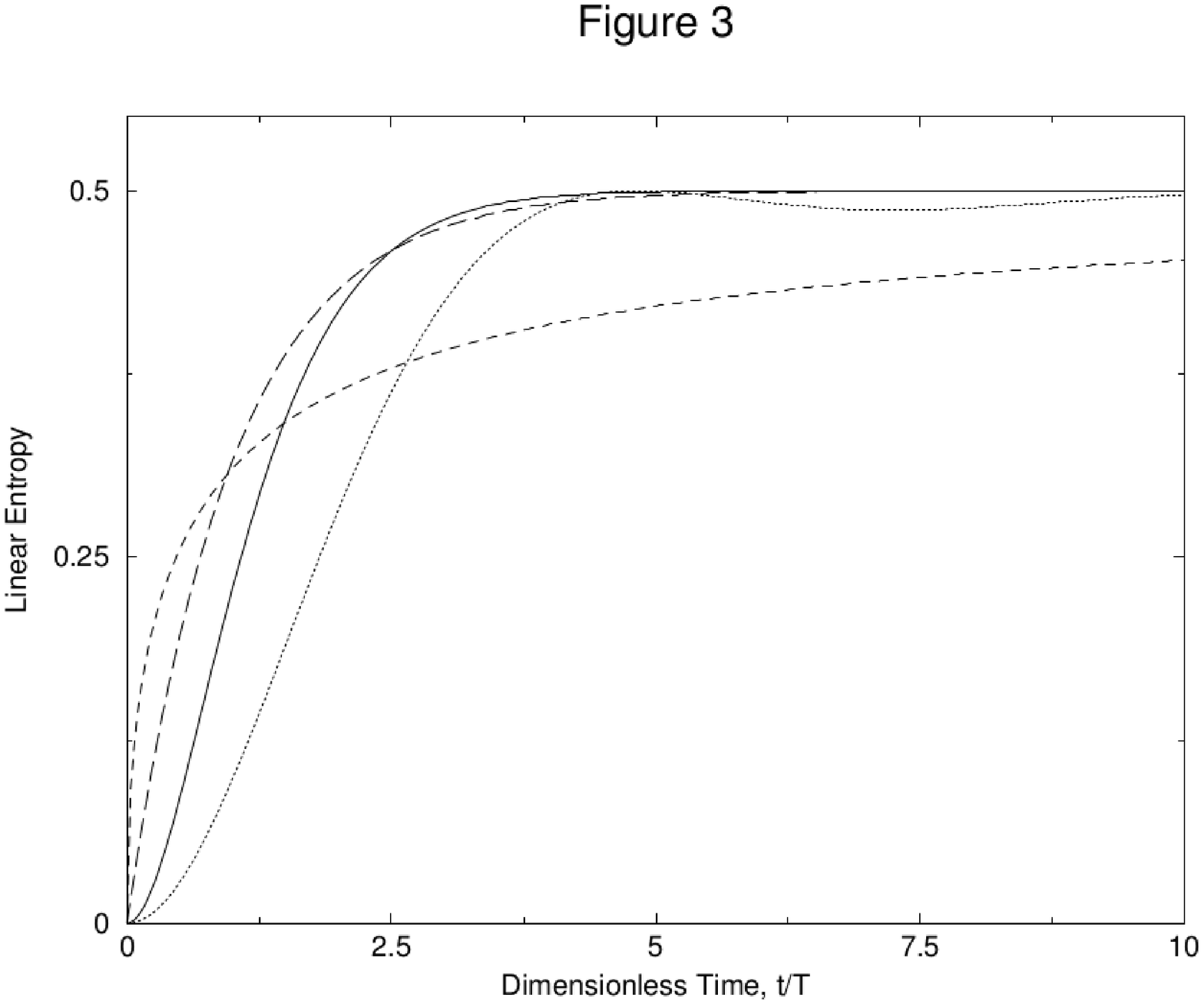}
\caption{Linear entropy  for the CTQRW defined by Eq.~(\ref{xy})
($p_{x}=p_{y}=1/2$). Long dashed line,
Markovian kernel with $A_{1}=0.5~sec^{-1}$. Dashed line, fractional kernel
with $\alpha=0.5$, $A_{\alpha}=1/\sqrt{2}~sec^{-1/2}$. Full line, exponential
kernel with $\gamma=2~sec^{-1}$, $A_{\epsilon}=1~sec^{-2}$. Dotted line,
exponential kernel with $\gamma=0.5~sec^{-1}$, $A_{\epsilon}=0.25~sec^{-2}$.}
\label{PositiveEntropy}
\end{figure}

In Fig.~(\ref{PositiveEntropy}) we show the linear entropy for the
Markovian, exponential and fractional kernels. As initial condition we have
chosen a pure state, an eigenstate of $\sigma_{x}$. In the case of the
exponential kernel, consistently, we  verify that independently of the
parameter values, the linear entropy is always
positive.

\subsubsection{Dephasing Reservoir}

Here, we assume that the superoperator ${\cal E}[\bullet]$ is defined through
the operator
\begin{equation}
C_{1}=\sigma _{z}.
\end{equation}
The Lindblad superoperator results in  ${\cal L}{[\bullet]}={\cal 
L}_{d}{[\bullet]}$,
where
\begin{equation}
{\cal L}_{d}\left[ \bullet \right] \equiv\frac{1}{2}([\sigma _{z},\bullet
\sigma _{z}]+[\sigma _{z}\bullet ,\sigma _{z}]).\label{disperso}
\end{equation}
As is well known, this kind of dispersive contribution destroys coherences
without affecting the level occupations.

In the case of the exponential kernel, the matrix elements are given by
\begin{equation}
P_{\pm }(t)=P_{\pm }(0),\;\;\;\;\;\;\;\;C_{\pm }(t)=C_{\pm }(0)h(t,\Phi _{d}).
\end{equation}
where the function $h(t,\Phi )$ was defined in Eq.~(\ref{dicotomico}) and
now $\Phi _{d}=\sqrt{\gamma^{2}-8 A_{\epsilon}}$.
It is simple to proof that
independently of any parameter value, here the evolution preserves the density matrix
positivity. This follows from the inequality
$d(t)=P_{+}(0)P_{-}(0)-C_{+}(0)C_{-}(0)(h(t,\Phi
_{d}))^{2}\geqslant 0$, which, added to the
preservation of the probability occupations,  guarantees
the positivity condition.
On the other hand, by expressing the density matrix in
the sum representation,
$\rho(t)=g_{I}(t) \rho(0)+g_{z}(t) \sigma_{z} \rho(0) \sigma_{z}$, with
$g_{I}(t)~=~[1+~h(t,\Phi_{d})]/2$ and $g_{z}(t)=[1-h(t,\Phi_{d})]/2$, it is
immediate to proof that the dynamics is completely positive for any parameter
values. Therefore, for this kind of dispersive superoperator,
independently of the possibility of associating to it a stochastic dynamics, the
solution map is always completely positive.

In the case of the fractional kernel we get
\begin{equation}
P_{\pm }(t)=P_{\pm }(0),\;\;\;\;\;\;\;C_{\pm }(t)=C_{\pm }(0)E_{\alpha
}[-2 A_{\alpha} t^{\alpha }].
\end{equation}
As in the previous environment model, here the coherence decay
displays stretched exponential and power law behaviors.

\subsubsection{Thermal Reservoir}

Now we will analyze a dynamics that leads to a thermal equilibrium state. First,
we assume
\begin{eqnarray*}
C_{1} &=&\sqrt{p_{\uparrow}}\left(
\begin{array}{cc}
1 & 0 \\
0 & \sqrt{1-\kappa}
\end{array}
\right), \;\;\;\;\;\;\;\;\;
C_{2} =\sqrt{p_{\uparrow}}\left(
\begin{array}{cc}
0 & \sqrt{\kappa} \\
0 & 0
\end{array}
\right), \\ C_{3}&=&\sqrt{p_{\downarrow}}\left(
\begin{array}{cc}
\sqrt{1-\kappa} & 0 \\
0 & 1
\end{array}
\right), \;\;\;\;\;\;\;\;\; C_{4}=\sqrt{p_{\downarrow}}\left(
\begin{array}{cc}
0 & 0 \\
\sqrt{\kappa} & 0
\end{array}
\right).
\end{eqnarray*}
where  $p_{\uparrow}+p_{\downarrow}=1$ and $0<\kappa \le 1$. These operators
correspond to a generalized amplitude damping superoperator \cite{nielsen}.
With these definitions, the Lindblad
superoperator Eq.~(\ref{lindblad}) can be written as
\begin{equation}
{\cal L}[\bullet]=\kappa {\cal L}_{th}[\bullet]+\tilde{\kappa}{\cal
L}_{d}[\bullet] \label{termico}
\end{equation}
where ${\cal L}_{d}[\bullet]$ was defined in Eq.~(\ref{disperso}), and
\begin{equation}
\tilde{\kappa}=\frac{1}{2}\left[1-\frac{\kappa}{2}-\sqrt{1-\kappa}\right]
\label{kapita}.
\end{equation}
On the other hand, the Lindblad term ${\cal L}_{th}[\bullet]$ corresponds to a
thermal reservoir
\begin{displaymath}
{\cal L}_{th}\left[ \bullet \right] \equiv\frac{ p_{\uparrow}}{2}([\sigma
^{\dagger
},\bullet \sigma ]+[\sigma ^{\dagger }\bullet ,\sigma ])+\frac{
p_{\downarrow} }{2} ([\sigma ,\bullet \sigma ^{\dagger }]+[\sigma \bullet
,\sigma
^{\dagger }]).
\end{displaymath}
The  temperature is defined by  $p_{\uparrow}/p_{\downarrow}=\exp[-\beta
\Delta E]$, where $\Delta E$ is the difference of energy between the two levels.

Before proceeding with the description of this case, we want to remark that
a pure thermal evolution  can be only introduced through an infinitesimal transformation.
 In fact, it is possible to demonstrate that the superoperator ${\cal E}_{th}[\bullet]\equiv {\cal
L}_{th}[\bullet]+\text {I}$ is not a completely positive one,
i.e., it can not be written in a sum representation Eq.~(\ref{super}). After noting that the
Lindblad superoperator Eq.~(\ref{termico}) satisfies
${\cal L}[\bullet ]=\kappa {\cal L}_{th}\left[ \bullet \right] + O(\kappa
^{2})$,
it is possible to  associate $\kappa$ with the control parameter  of Eq.~(\ref{infinitesimal}). 
Thus, in the limit $\kappa \rightarrow 0$ the dispersive contribution drops out.

The dynamics induced by the Lindblad Eq.~(\ref{termico}) is similar to those
analyzed previously in this section.
In fact, the solution for the exponential case can be written as in
Eqs.~(\ref{popular})-(\ref{coherente}) with
\begin{equation}
\Phi _{pop} =\sqrt{\gamma^{2}-4  \kappa A_{\epsilon} } ,\;\;\;\;
\\  \Phi _{coh} =\sqrt{\gamma^{2}-2 (\kappa+4 {\tilde \kappa})
A_{\epsilon}}.\label{rateexponential}
\end{equation}
On the other hand, for the fractional kernel, the solutions read as in
Eqs.~(\ref{popfrac})-(\ref{coherfrac}) with the definitions
\begin{equation}
\Phi_{pop}^{(\alpha)}=\kappa A_{\alpha},\;\;\;\;\;\;
\Phi_{coh}^{(\alpha)}=(\frac{\kappa}{2}+2{\tilde
\kappa})A_{\alpha}.   \label{ratefractional}
\end{equation}

The main difference with the previous solutions are the equilibrium
populations which now read $P_{+}^{eq}=p_{\uparrow}$, and
$P_{-}^{eq}=p_{\downarrow}$.
As a consequence of this fact, it is simple to realize that for $\gamma^{2}
\le A_{\epsilon}$, the exponential kernel produces a mapping that is not
completely positive and not even positive. This follows by noting that for
$P_{\pm}^{eq}\ne 1/2$, the population solutions Eq.~(\ref{popular}) can take
negative values.

\begin{figure}[tbp]
\includegraphics[bb=94 75 709 581,angle=0,width=8.cm]{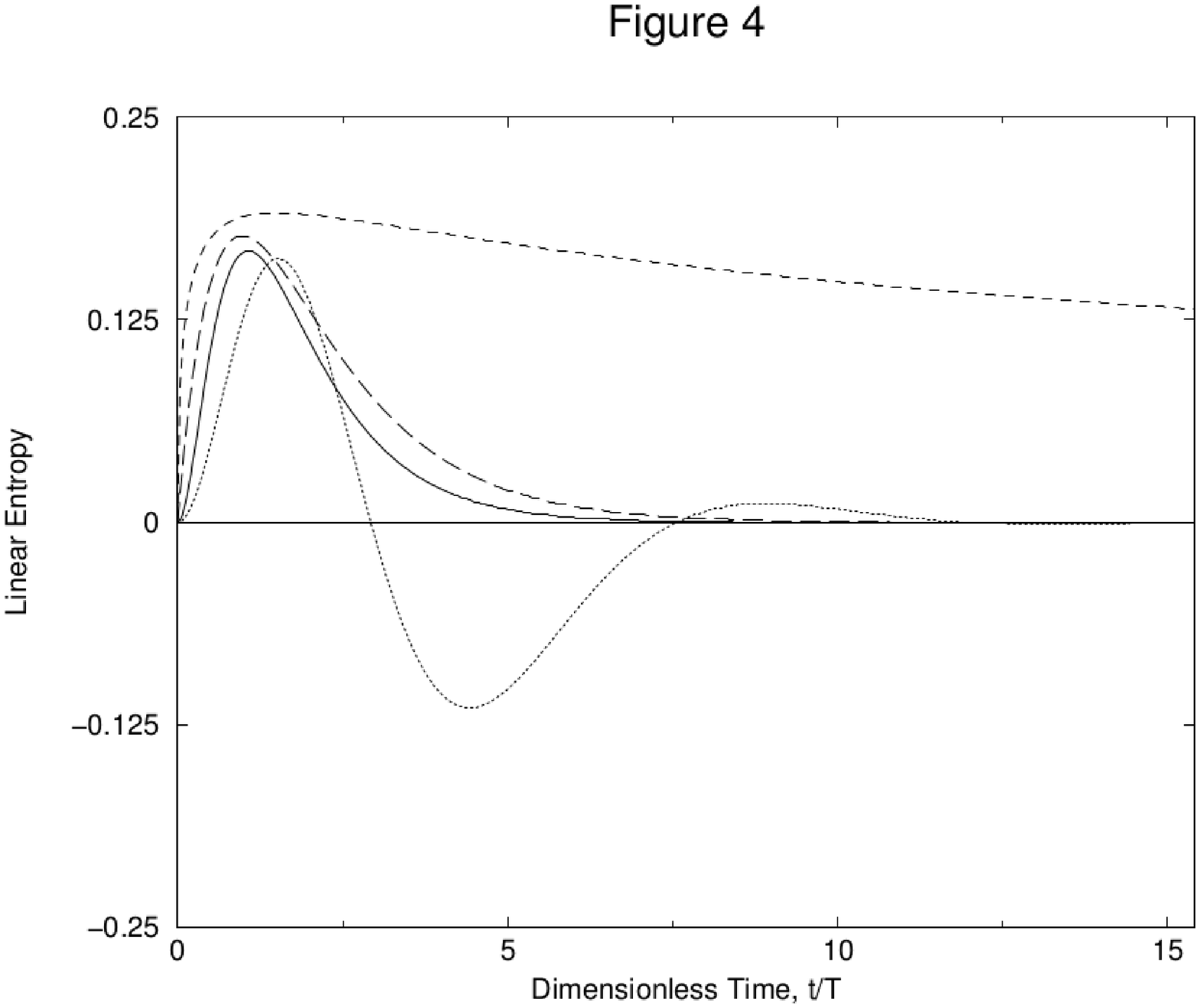}
\caption{Linear entropy  for the CTQRW defined by Eq.~(\ref{termico})
with $p_{\downarrow}=1$, $p_{\uparrow}=0$, and $\kappa=0.75$. Long dashed Line,
Markovian kernel with $A_{1}=1~sec^{-1}$. Dashed line, fractional kernel
with $\alpha=0.5$, $A_{\alpha}=1~sec^{-1/2}$. Full line, exponential kernel
with $\gamma=4~sec^{-1}$, $A_{\epsilon}=4~sec^{-2}$. Dotted line,
exponential kernel with $\gamma=1~sec^{-1}$, $A_{\epsilon}=1~sec^{-2}$.}
\label{NegativeEntropy}
\end{figure}

In Fig.~(\ref{NegativeEntropy}), for each kernel, we show the linear entropy
behavior in the case of a zero temperature reservoir. As in the previous figure,
as initial condition we use an eigenstate of the $x-$Pauli matrix.
In the exponential case, when the stochastic interpretation is not possible the
linear entropy takes negative values. Equivalently, this means that
$\rho(t)$ is not positive definite.


\subsection{Dynamics in a Fock Space}

Here we will analyze the dynamics of a CTQRW in a system
provided with a Fock space structure, as for example a quantum harmonic
oscillator or a mode of an electromagnetic field. With $a^{\dagger}$ and $a$ we
denote the corresponding creation and anhilation operators. This situation
will allow us to recover the classical concept of continuous time random walks
in the context of completely positive maps.

For the superoperator that defines the CTQRW, we assume the following form
\begin{equation}
{\cal E}[\rho]=D_{(\beta,\beta^{*})} \rho D_{(\beta,\beta^{*})}^{\dagger},
\end{equation}
where $D_{(\beta,\beta^{*})}$ is the displacement operator
\begin{equation}
D_{(\beta,\beta^{*})}=\exp[\beta a^{\dagger}-\beta^{*} a].
\end{equation}
Furthermore, we assume that in each application of ${\cal E}[\bullet]$ the
complex parameter $\beta$ is chosen with a probability distribution
$P_{(\beta,\beta^{*})}$.
The induced evolution can be easily analyzed by introducing the Wigner
function
\begin{equation}
W( \alpha,\alpha^{*},t) =2 Tr [\rho(t)  D_{(\alpha,\alpha^{*})}
e^{i \pi a^{\dagger} a}
D_{(\alpha,\alpha^{*})}^{\dagger}],
\end{equation}
whose evolution from Eqs.~(\ref{master})-(\ref{LA}) then reads
\begin{eqnarray}
\frac{d W(\alpha,\alpha^{*},t)}{dt}&=&\int_{0}^{t} d\tau K(t-\tau)
\bigg\{\int_{-\infty}^{\infty}
 d\beta d\beta^{*}  P_{(\beta,\beta^{*})} \label{eugenia}
\\ &&
  W(\alpha-\beta,\alpha^{*}-\beta^{*},\tau)-
W(\alpha,\alpha^{*},\tau)\bigg\}.\nonumber
\end{eqnarray}
By construction, the solution of this equation provides a completely positive
map. Furthermore, we note that this equation can be interpreted as a ``classical'' continuous time random walk where the statistic of the ``particle jumps'' is given by   $P_{(\beta,\beta^{*})}$ and the statistics of the elapsed time between the successive jumps is characterized through the waiting time distribution associated to the kernel $K(t)$. Thus, it is evident that this evolution is a classical
one \cite{fischer}, which implies that any quantum property can only be introduced through the initial conditions.

When  all the moments of the distribution
$P_{(\beta,\beta^{*})}$ are finite, i.e., $\langle \beta^{r} \beta^{* s}
\rangle \equiv \int_{-\infty}^{\infty}
 d\beta d\beta^{*}  P_{(\beta,\beta^{*})} \beta^{r} \beta^{* s}<\infty$
$\forall$ $r$, $s$, the evolution Eq.~(\ref{eugenia}) can be written in terms of a Kramers-Moyal expansion
\begin{equation}
\frac{d W(\alpha,\alpha^{*},t)}{dt}=\int_{0}^{t} d\tau K(t-\tau)
{\cal L} W(\alpha,\alpha^{*},\tau), \label{kramer}
\end{equation}
where the operator ${\cal L}$ is defined by
\begin{equation}
{\cal L}=\sum_{n=1}^{\infty} \frac{1}{n !} \int_{-\infty}^{\infty}
 d\beta d\beta^{*}  P_{(\beta,\beta^{*})} \left(\beta \frac{\partial}{\partial
\alpha} +\beta^{*}
\frac{\partial}{\partial \alpha^{*}}\right)^{n}.
\end{equation}
These expressions follow after developing in Eq.~(\ref{eugenia}) 
 the Wigner function $W(\alpha-\beta,\alpha^{*}-\beta^{*},\tau)$  around 
$W(\alpha,\alpha^{*},\tau)$.
In this situation, it is also possible to get a close expression for the average excitation number
$n(t)=Tr[\rho(t) a^{\dagger}a]$, which reads
\begin{equation}
n(t) =n(0)+\langle |\beta|^{2} \rangle \int_{0}^{t}d\tau   K(t-\tau)
\tau.\label{excitacion}
\end{equation}
Here, we have assumed that the average displacements in the directions $(\beta, \beta^{*})$ are null, i.e., the first moments of the distribution $P_{(\beta,\beta^{*})} $ vanish.

Up to second order, the operator ${\cal L}$ reduces to a Hamiltonian term plus a
classical Fokker Planck operator. By truncating the evolution up to this order,  
the CPC  is not broken.
This fact can be easily demonstrated by going back to the
density matrix representation, where the Lindblad superoperator
Eq.~(\ref{LA}) then reads
${\cal L}[\bullet] \approx {\cal L}_{H}[\bullet]+{\cal L}_{FP}[\bullet]$, with
\begin{equation}
{\cal L}_{H}\left[ \bullet \right] =[\langle \beta \rangle a^{\dagger}-\langle
\beta^{*} \rangle a,\bullet], \label{shift}
\end{equation}
and
\begin{eqnarray}
{\cal L}_{FP}\left[ \bullet \right] &=& \langle |\beta|^{2} \rangle([a^{\dagger
},\bullet
a]+[a^{\dagger }\bullet ,a]+[a,\bullet a^{\dagger
}]+[a\bullet ,a^{\dagger }]) \nonumber
\\ &+& \langle \beta^{2} \rangle([a^{\dagger
},\bullet
a^{\dagger}]+[a^{\dagger }\bullet ,a^{\dagger}]) \nonumber
\\ &+& \langle \beta^{* 2} \rangle([a,\bullet
a]+[a\bullet ,a]). \label{secondLindblad}
\end{eqnarray}
The Lindblad terms proportional to $\langle |\beta|^{2} \rangle$ are
equivalent to a reservoir at infinite temperature and the terms proportional to
$\langle \beta^{2} \rangle$ and $\langle \beta^{*2} \rangle$
introduce a squeezing effect.
On the other hand, it is possible to demonstrate that maintaining only a finite number of higher
terms, the evolution for the density matrix can not be written in a
Lindblad form and in consequence it is not completely positive. This fact
agrees with the predictions of the classical Pawula theorem \cite{pawula} about
Fokker Planck equations.

{\it Subdiffusive Processes}: By assuming the fractional kernel Eq.~(\ref{kernelfrac}), in the 
limit $A_{\alpha} \rightarrow \infty$,  $\langle |\beta|^{2} \rangle \rightarrow 0$, with 
$A_{\alpha} \langle |\beta|^{2} \rangle= A_{\alpha}^{'}$, the previous second order approximation 
applies. In  this situation, the evolution of the Wigner function is characterized by a subdiffusive process. In fact, the average excitation number reads
\begin{equation}
n(t) =n(0)+\frac{2  A_{\alpha }^{'}}{\Gamma (1+\alpha )}
t^{\alpha}.
\end{equation}
Note that  in comparison with a Markovian Lindblad evolution, $\alpha=1$, here the increasing of the average excitations present a slower grow. On the other hand, the evolution of the Wigner function can be written as
\begin{equation}
\frac{\partial W\left( x,t\right) }{\partial t}= A_{\alpha}^{'}\;_{0}D_{t}^{1-\alpha }
\frac{\partial ^{2}}{\partial x^{2}}W\left( x,t \right). \label{equis}
\end{equation}
Here, $x$ is an arbitrary direction in the complex plane, and in order to simplify the expression, we have ``traced out'' the Wigner function over the perpendicular direction.
We remark that this kind of fractional subdiffusive dynamics is allowed in the context of completely positive maps.
This equation was extensively analyzed in the literature \cite{metzler}, 
where it was found  that the solution presents a  non-Gaussian
diffusion front. We notice that the relations between the
exponents that characterize this behavior \cite{front} were found to be universal in the
context of quasiperiodic and disordered systems \cite{zhong}.


{\it Long Jumps}:
When the moments of the distribution
$P_{(\beta,\beta^{*})}$ are not defined, the dynamics must be analyzed in the Fourier domain,
 $(\alpha,\alpha^{*})\rightarrow (k,k^{*})$. Denoting with a hat symbol the Fourier
transform, from Eq.~(\ref{eugenia}), we get
\begin{displaymath}
\frac{d \hat{W}(k,k^{*},t)}{dt}=-\gamma(k,k^{*}) \int_{0}^{t} d\tau K(t-\tau)
\hat{W}(k,k^{*},\tau),
\end{displaymath}
where the rates of the Fourier modes is given by
\begin{equation}
\gamma(k,k^{*})=1-\hat{P}_{(k,k^{*})}.
\end{equation}
For example, by assuming a Levy distribution \cite{metzler}
$P_{(k,k^{*})}=exp[-\sigma^{\mu}|k|^{\mu}]$, with $0<\mu \le 2$,
 the evolution can be written as a series
of infinite fractional derivatives with respect to the variables
$(\alpha,\alpha^{*})$. Nevertheless, with the present formalism, it is not
possible to check the CPC of any truncated evolution.


{\it Quantum Random Walks}: Finally we note that the concept of quantum
random walks \cite{julia} used in the context of quantum computation and quantum
information can be recovered as a particular case of our approach by using the
generalized displacement operator
\begin{equation}
D_{(\beta,\beta^{*},\theta,\phi)}=R(\theta,\phi) \;exp[\sigma_{z}(\beta
a^{\dagger}-\beta^{*}
a)],
\end{equation}
and assuming that
$P_{(\beta,\beta^{*})}=\delta_{(\beta-\beta_{0})}\delta_{(\beta^{*}-\beta_{0}^{*})}$, and
$w(t)=\delta(t-T_{0})$. Here, $R(\theta,\phi)$ is
an arbitrary rotation of an extra spin variable,  $(\beta_{0},\beta_{0}^{*})$ is
an arbitrary direction in the complex plane and $T_{0}$ is the discreet time
step.

\subsection{Generalized Intrinsic Decoherence Formalism}

The intrinsic decoherence formalism \cite{milburn,moya} was introduced by Milburn as a
phenomenological frame to the description of decoherence phenomema. Here, we
will analyze and generalize this formalism by interpreting it as a CTQRW. First,
we assume as a superoperator
\begin{equation}
{\cal E}_{\tau }[\bullet ]=e^{-iH\tau }\bullet e^{iH\tau},
\end{equation}
where $H$ is an arbitrary Hamiltonian in a given Hilbert space, and $\tau$ is a
random variable chosen with a density probability $P(\tau)$.
From Eqs.~(\ref{master})-(\ref{LA}), the average density matrix evolves as
\begin{eqnarray}
\frac{d\rho (t)}{dt}=\int_{0}^{t} d\tau K(t-\tau)
\bigg\{&&\int_{-\infty }^{+\infty} d\tau^{'}  P(\tau^{'} )
\\ && e^{-iH\tau^{'} }\rho (\tau)e^{iH\tau^{'} }-\rho (\tau) \bigg\}.\nonumber
\end{eqnarray}
In the basis of eigenstates of the Hamiltonian $H$, $H
|n\rangle=\varepsilon_{n}|n\rangle$, the evolution of the matrix elements
$\rho_{nm}=\langle n|\rho | m \rangle$ is given by
\begin{equation}
\frac{d\rho _{nm}(t)}{dt}=-\gamma _{nm} \int_{0}^{t} d\tau K(t-\tau) \rho
_{nm}(\tau).
\end{equation}
Here, the decaying rates $\gamma_{nm}$ read
\begin{equation}
\gamma _{nm}=1-\hat{P}(\omega_{nm}),
\end{equation}
where $\hat{P}(\omega)=\int_{-\infty}^{\infty} d\tau P(\tau)e^{-i \omega\tau}$,
is  the Fourier transform of the probability and
$\omega_{nm}=\varepsilon_n-\varepsilon_m$ are the Bohr frequencies.

The original Milburn proposal is obtained by choosing
\begin{equation}
w(t)=(1/\tau _{a})\exp (- t/\tau _{a}),\;\;\;\;\;P(\tau)=\delta (\tau -\tau
_{b}),\label{intrinseco}
\end{equation}
which implies the density matrix evolution
\begin{equation}
\frac{d\rho (t)}{dt}=\frac{1}{\tau _{a}}\left\{ e^{-iH\tau _{b}}\rho
(t)e^{iH\tau _{b}}-\rho (t)\right\}.
\end{equation}
Thus, our CTQRW provides a natural non-Markovian generalization of this
formalism. On the other hand, by choosing the  exponential waiting distribution
of Eq.~(\ref{intrinseco}),
$P(\tau)=(t/\tau_{b})^{-1}\exp (-t/\tau_{b})$, and using the
identity $\ln (1+ix)=\int_{0}^{\infty }ds(e^{-s}/s)(1-e^{isx})$, the rate
results $\gamma _{nm}=\ln (1+\omega _{nm}\tau _{b})]/\tau _{a}$. This expression
coincides with that obtained in the formalism of
Ref.~\cite{bonifaccio}.


\section{Summary and Conclusions}

In this paper we have demonstrated that  non-Markovian master equations that
consist in a  memory integral over a Lindblad structure can be considered
as a valid tool in the description of open quantum system dynamics.

Our approach for the understanding of this kind of equations
consists in a natural generalization of the classical concept of continuous
time random walks to a quantum context.
We have defined a CTQRW in terms of a set of random renewal
events, each one consisting in the action of a superoperator over a
density matrix.
The selection of different statistics for the elapsed time between the successive
applications of the superoperator allowed us to construct different classes of
completely positive evolutions that lead to strong non-exponential decay of the
density matrix elements. Remarkable
examples are the telegraphic master equation, Eq.~(\ref{telegrafo}), which
interpolates between a Gaussian short time dynamics and an asymptotic exponential
decay, and the fractional master equation, Eq.~(\ref{fraccionaria}), which leads
to stretched exponential and power law behaviors.
On the other hand, in a Fock space the dynamics reduces to a classical one, 
which allowed us to demonstrate that
fractional subdiffusive processes are  consistent with a completely positive evolution.

Concerning the possibility of obtaining non-physical solutions
from the Non-Markovian master equation Eq.~(\ref{master}), we have found a set
of mathematical conditions on the kernel
that guarantee the CPC of the
solution map. As in classical Fokker-Planck equations, the set of conditions
Eq.~(\ref{condiciones}) allows us to link each safe kernel with a
corresponding waiting time distribution, which in the present case allows
to associate to the master equation a CTQRW.

By analyzing the exponential kernel, related to the telegraphic master
equation, we have demonstrated that when the kernel can not be associated with a waiting time distribution, the resulting solution map can be either non-physical,
only positive, or even completely positive.
This case demonstrates that no general conclusions can be obtained outside the
regime where a stochastic interpretation is available.
Furthermore, we have demonstrated that telegraphic master equations constructed 
with  Lindblad
superoperators that can be only introduced through an infinitesimal
transformation,
 Eq.~(\ref{infinitesimal}), only admit a stochastic interpretation in the 
Markovian limit.
In the case of the fractional kernel we have implemented a numerical simulation that
confirms the equivalence between the  non-Markovian fractional master equation  and the
corresponding  CTQRW.

Finally we want to remark that from the understating achieved in this work,
some interesting open question arise in a natural way, as for example a possible microscopic derivation of these non-Markovian master equations and the  finding of alternative  stochastic representation based in a continuous measurement theory. 
In fact, from the examples worked out in  this paper,  we
conclude that the stochastic dynamics of a CTQRW can be thought in a rough way
as the continuous measuring action of an environment over
an open quantum system, where the scattering superoperator must be
associated with the microscopic interaction
between the system and the environment, and the statistics of the random times
with the spectral properties of the bath.

\acknowledgments
I am grateful to  H. Schomerus and D. Spehner for enlighting discussions.


\begin{references}

\bibitem{alicki}  R. Alicki and K. Lendi, in {\it Quantum Dynamical
Semigroups and Applications}, (Lect. N. in Phys. {\bf 286}, Springer, 1987).

\bibitem{nielsen} M.A. Nielsen and I.L. Chuang, {\it Quantum Computation and
Quantum Information}, Cambridge University Press (2000).


\bibitem{gruebele} V. Wong and M. Gruebele, Chem. Phys. {\bf 284}, 29 (2002).

\bibitem{wong} V. Wong and M. Gruebele, Phys. Rev. A {\bf 63}, 022502
(2001).

\bibitem{zhong} J. Zhong, {\it et. al.}, Phys. Rev. Lett. {\bf 86}, 2485 (2001).

\bibitem{barkai_dot}  Y. Jung, E. Barkai, and R.J. Silbey, Chem. Phys. {\bf
284}, 181 (2002).

\bibitem{dobro} V.V. Dobrovitski, {\it et. al.}, Phys. Rev. Lett. {\bf 90}, 210401
(2003).


\bibitem{kuznezov}  D. Kuznezov, A. Bulgac, and G.D. Dang, Phys. Rev. Lett.
{\bf 82}, 1136 (1999).


\bibitem{dalton} B.J. Dalton and B.M. Garraway, Phys. Rev. A {\bf 68}, 033809
(2003).


\bibitem{feller}  W. Feller, {\it An introduction to Probability Theory and
Its Applications} (J. Wiley \& Sons, NY, 1971), Vols. 1 and 2.


\bibitem{montroll} E.W. Montroll and G.H. Weiss, J. Math. Phys. {\bf 6}, 167
(1965); H. Scher and E.W. Montroll, Phys. Rev. {\bf 12}, 2455 (1975).

\bibitem{metzler}  R. Metzler, and J. Klafter, Phys. Rep. {\bf 339}, 1
(2000).


\bibitem{julia} J. Kempe, arXiv: quant-ph/0303081 (2003).



\bibitem{barnett}  S.M. Barnett, and S. Stenholm, Phys. Rev. A {\bf 64},
033808 (2001).



\bibitem{sokolov}  I.M. Sokolov, Phys. Rev. E {\bf 66}, 041101 (2002).

\bibitem{barkai} R. Metzler, E. Barkai, and J. Klafter, Phys. Rev. Lett. {\bf
82}, 3563 (1999).


\bibitem{aclaracion} In Ref.\cite{sokolov} the extra conditions
$\tilde{K}\left( u\right) \geqslant 0$
and that $(d/du)\tilde{K}(u)$ must be a CM function are also demanded. By writing the survival probability as
\[{\tilde P}_{0}(u)=\frac{1}{{\tilde K}(u)[1+u/{\tilde K}(u)]},
\] it is simple to realize that these conditions are equivalent to demand the inequality
$P_{0}(t) \ge 0$,  which implies that this extra conditions are automatically satisfied
if the conditions Eq.~(\ref{condiciones}) are satisfied.
This follows by noticing that a well defined waiting time distribution always  guarantee
$P_{0}(t)\ge 0$.


\bibitem{briegel}  H.J. Briegel, and B.G. Englert, Phys. Rev. A {\bf 47},
3311 (1993).


\bibitem{kimura} G. Kimura, Phys. Rev. A {\bf 66}, 062113 (2002).


\bibitem{morse}  P.M. Morse and H. Feshbach, {\it Methods of Theoretical
Physics}, Mc Graw Hill Company, (1953).


\bibitem{lu} Lu-Ming Duan and Guang-Can Guo, Phys. Rev. A {\bf 56}, 4466 (1997).

\bibitem{budini}  A.A. Budini, Phys. Rev. A {\bf 64}, 052110 (2001).


\bibitem{yu} S. Yu, Phys. Rev. A {\bf 62}, 024302 (2000).


\bibitem{fischer} W. Fischer, H. Leschke, and P. M\"{u}ller, Phys. Rev. Lett.
{\bf 73} 1578 (1994).


\bibitem{pawula} R.F. Pawula, Phys. Rev. {\bf 162}, 186 (1967).




\bibitem{front} See Eq.~(45) in Ref.~\cite{metzler}.



\bibitem{milburn}  G.J. Milburn, Phys. Rev. A {\bf 44}, 5401 (1991).
\bibitem{moya}  H. Moya-Cessa, V. Bu\v{z}ek, M.S. Kim, and P.L. Knight,
Phys. Rev. A {\bf 48}, 3900 (1993).


\bibitem{bonifaccio}  R. Bonifacio, {\it et. al.}, J. Mod. Opt. {\bf 47}, 2199
(2000);  R. Bonifacio, S. Olivares, P. Tombesi, and D. Vitali,
Phys. Rev. A {\bf 61}, 053802 (2000).

\end{references}
\end{document}